\documentclass[runningheads]{llncs} 

\usepackage{booktabs} % For formal tables
\usepackage{./why3lang}
\usepackage{listings}

\title{CISE3: Verifying Weakly Consistent Applications with Why3}

\author{
Filipe Meirim\and
M\'ario Pereira\and
Carla Ferreira}

\authorrunning{F.\ Meirim et al.}

\institute{ NOVA LINCS, FCT, 
Universidade Nova de Lisboa, Portugal%
}

\begin{document}

\maketitle

\begin{abstract}
  In this paper we present a tool for the formal analysis of applications built
  on top of replicated databases, where data integrity can be at stake. To
  address this issue, one can introduce synchronization in the
  system. Introducing synchronization in too many places can hurt the system's
  availability but if introduced in too few places, then data integrity can be
  compromised. The goal of our tool is to aid the programmer reason about the
  correct balance of synchronization in the system. Our tool analyses a
  sequential specification and deduces which operations require synchronization
  in order for the program to safely execute in a distributed environment. Our
  prototype is built on top of the deductive verification platform Why3, which
  provides a friendly and integrated user experience. Several case studies have
  been successfully verified using our tool.
\end{abstract}

%\begin{CCSXML}
%<ccs2012>
%  <concept_id>10010520.10010553.10010562</concept_id>
%  <concept_desc>Computer systems organization~Embedded systems</concept_desc>
%  <concept_significance>500</concept_significance>
%  <concept_id>10010520.10010575.10010755</concept_id>
%  <concept_desc>Computer systems organization~Redundancy</concept_desc>
%  <concept_significance>300</concept_significance>
%  <concept_id>10010520.10010553.10010554</concept_id>
%  <concept_desc>Computer systems organization~Robotics</concept_desc>
%  <concept_significance>100</concept_significance>
%  <concept_id>10003033.10003083.10003095</concept_id>
%  <concept_desc>Networks~Network reliability</concept_desc>
%  <concept_significance>100</concept_significance>
%</ccs2012>
%\end{CCSXML}

%\pagestyle{plain}
%
%\keywords{Why3, Data replication, Software Verification, Concurrency, \newline Weak Consistency}

\section{Introduction}
\label{intro}
Nowadays most large-scale distributed applications  depend on  geo-replicated storage systems in order to improve user experience. Geo-replicated storage systems consist of several replicas scattered around the world storing copies of an application's data. In this kind of storage systems, requests are routed to the nearest replica, making the system more available to the user. However, when updates occur simultaneously over different replicas, data integrity can be compromised. A possible solution for this issue would be to introduce synchronization in the system. If the programmer decides to employ a strong consistency model, where a total order of the execution of operations in different replicas is guaranteed, data integrity is ensured but the availability of the system decreases. On the other hand, if a weak consistency model is employed, where a total order of operations in different replicas is not guaranteed, the system becomes more available but data integrity might be broken.

In order to achieve a balance between strong and weak consistency models, it was proposed that geo-replicated systems could use a combination of these models \cite{balegas2015,balegas2016}. Specifically, the approach would be to use strong consistency whenever the correction of the application is at risk, and employ weak consistency when concurrent execution is safe. However, finding a balance between strong and weak consistency, in order to maximise the application's availability, is a non-trivial task \cite{CAP}. The programmer needs to reason about the concurrent effects of operations, and decide which operations require synchronization to assure the correctness of the application.

In this paper, we propose an automatic approach to analyse weakly consistent distributed applications using the deductive verification platform Why3 \cite{filliatre:hal-00789533} and following the proof rule presented by the CISE logic \cite{cise}. The goal of the proof rule is to identify which pairs of operations cannot be safely executed concurrently, and consequently require synchronization. Throughout this paper we illustrate our approach using a complete case study written and verified with our tool.

This paper is organized as follows. In Section \ref{cise} we present a brief overview of the CISE logic and its underlying proof rule. In Section \ref{why3} we briefly overview the Why3 framework. In Section \ref{cise3} we show our approach and demonstrate how it works. In Section \ref{crdts} we present our approach for the resolution of commutativity issues. In Section \ref{case}, we illustrate how our tool works by means of a complete case study. In Section \ref{related} we present some related work. We conclude this paper in Section \ref{conclusion}, also discussing possible future work.

\section{Preliminary concepts}
\label{cise}

The CISE logic \cite{cise} presents a proof rule for analysing the preservation of integrity invariants in applications built over replicated databases. This logic presents the concept of a generic consistency model where a wide range of consistency models can be expressed, like weak consistency or strong consistency, for each operation of the application (e.g., Parallel Snapshor Isolation (PSI) and explicit consistency).

A consistency model is expressed via a token system, containing a set of tokens $T$ and a conflict relation $\bowtie$ over $T$. Each operation of the system has associated a (possibly empty) set of tokens, that it must acquire so it can execute. Operations with conflicting tokens cannot be executed concurrently and require synchronization. However, if the set of tokens associated to an operation is empty, then that operation does not require any synchronization and weak consistency can be used.

For example, the protocol of mutual exclusion can be expressed with a single token $\tau$ conflicting with itself ($\tau \bowtie \tau$). If operations are accessing a shared resource, then they should be associated to token~$\tau$. This way, it is guaranteed that when an operation acquires the token $\tau$, no other operation trying to access the resource can be executed concurrently. Therefore, mutual exclusion is guaranteed.

The CISE proof rule allows two different verification approaches. The first approach consists in verifying the validity of a token system previously defined by the programmer. The second approach consists in identifying the pairs of conflicting operations that can break the application's invariants. These conflicting operations already define an initial coarse-grained token system, that can then be refined by the programmer.

In operational terms, the proof rule is composed by the following analyses:

\begin{itemize}
	\item \textbf{Safety analysis:} verifies if the effects of an operation when executed without any concurrency, preserve the invariants of the application.
	\item \textbf{Commutativity analysis:} verifies if every pair of (different) operations commute, \textit{i.e.}, if alternative execution orders reach the same final state, starting from the same initial state.
	\item \textbf{Stability analysis:} verifies if the preconditions of each operation are stable under the effects of all other operations of the system.
\end{itemize}

The proof rule of CISE was automatised with the purpose of reasoning about the correction of distributed applications executed over weakly consistent databases \cite{Marcelino:2017:BHC:3064889.3064896,Najafzadeh:2016:CTP:2911151.2911160}. The automatisation uses SMT solvers in order to analyse the underlying verification conditions.

\section{Why3 overview}
\label{why3}

Why3 is a framework used for the deductive verification of programs, \textit{i.e.}, "the process of turning the correctness of a program into a mathematical statement and then proving it" \cite{memoire}. Why3's architecture is divided in two components: a purely logical back-end and a programs front-end \cite{memoire}. The front-end part receives as input, files that contain a list of modules from which verification conditions will be extracted, and subsequently sent to external theorem demonstrators.

This tool provides a programming language called WhyML, which has two purposes: writing programs and the formal specifications of their behaviours. WhyML is a first order language with some features frequently found in functional languages, like, pattern-matching, algebraic types and polymorphism. Simultaneously, WhyML has some imperative traits like records with mutable fields and exceptions. The logic used to write the specifications is an extension of first order logic with polymorphic types, algebraic types, inductive predicates, recursive definitions, as well as a limited form of higher order logic \cite{modular_way}. Another useful trait of the WhyML language is ghost code, which is used to write the program's specifications and aid the proof of said program. A particularity of ghost code is that it can be removed from a program without affecting its execution. This happens because ghost code cannot be used in a regular code computation, and cannot change a mutable value of regular code \cite{ghost_code}. Also, regular code cannot modify nor access ghost code.

The WhyML language can also be used as an intermediate language for the verification of programs written in C, Ada, or Java for example \cite{filliatre:hal-00789533}. These programs can be translated into a WhyML program, where finally the verification conditions are extracted and sent to external provers. The tools Frama-C \cite{Kirchner:2015:FSA:2769048.2769092}, Spark2014 \cite{spark}, and Krakatoa \cite{krakatoa} use WhyML as an intermediate language for the verification of programs written in C, Ada, and Java respectively.

Simultaneously, the Why3 framework also provides a graphic environment for the development of programs, where it is possible to interact with several theorem provers, perform small interactive proof steps \cite{dailler2018}, as well as, visualising counter-examples.

The Why3 platform serves as a front-end to communicate with more than 25 interactive and automated theorem provers. This framework has already been used to prove several realistic programs including an OCaml library (VOCaL) \cite{parreirapereira:tel-01980343}, a certified first order theorem prover \cite{clochard:hal-00913431}, an interpreter for the CoLiS language \cite{colis}, the Strassen's algorithm for matrix's multiplication \cite{Clochard2018}, and an efficient arbitrary-precision integer library \cite{rieuhelft17vstte}.

\section{CISE3}
\label{cise3}

The Why3 framework can be extended by plug-ins like Jessie \cite{jessie} and Krakatoa \cite{krakatoa}. The integration of new plug-ins into Why3 is relatively simple: write a parser for the target language, whose intermediate representation should be mapped to a non-typed AST of the WhyML language. The CISE3 tool is a plug-in for the Why3 framework. In our particular case, we used the already existing parser for the WhyML language. We believe that our choice of the Why3 framework, and basing our tool on its architecture of plug-ins, leads to a faster development and a reduction of the validation effort. On the other hand, developing our tool over a mature framework allows us to scale up for the analysis of more realistic examples. In this section we describe how the CISE3 tool works and we illustrate each of the conducted CISE analyses.

\subsection{Safety analysis}
\label{safety}

This analysis consists in verifying if an operation, when executed without any concurrency, can break an integrity invariant of the application. The programmer needs to provide as input to the tool the specification of the application state and its invariants, as well as the sequential implementation and specification of each operation. Let's consider, as an example, the generic program in Figure \ref{generic_prog} composed by operations \of{f} and \of{g}, and the type of the state of the application $\tau$ with the invariant $\mathcal{I}$ associated to it.

\begin{figure}
\begin{why3} [label = generic]
  			type $\tau$[@state] = { $\overline{\mathtt{x:\tau_x}}$ }
  			invariant { $\mathcal{I}$ }

  			let f ($\overline{\mathtt{x:\tau_1}}$) (state: $\tau$)
    			  requires { $\mathcal{P}_1$ }
    			  ensures  { $\mathcal{Q}_1$ }
  			= e$_1$

  			let g ($\overline{\mathtt{y:\tau_2}}$) (state: $\tau$)
    			  requires { $\mathcal{P}_2$ }
    			  ensures  { $\mathcal{Q}_2$ }
  			= e$_2$
	\end{why3}
	\caption{Generic Why3 program.}
	\label{generic_prog}
\end{figure}

The programmer must associate the tag \of{[@state]} with the specification of the state type, so our tool can identify it as the application state, as seen in Figure \ref{generic_prog}. The type $\tau$ possesses a set of fields represented by $\overline{\texttt{x}}$, whose types are represented as $\overline{\tau_{\texttt{x}}}$. Additionally, Figure \ref{generic_prog} presents the definition of operations \of{f} and \of{g}. Every operation of the application must have an instance of the state of the application passed as a parameter. This is used later in the generation of commutativity and stability analysis functions. The preconditions of operation \of{f} are represented as $\mathcal{P}_1$, the postconditions are represented as $\mathcal{Q}_1$, and its body is represented as \of{e}$_1$. The specification of operation \of{g} is similar to the one of \of{f}.

The Why3 framework by itself is capable of verifying the safety of each operation. This amounts to a traditional proof to check that the implementation adheres to the supplied specification. Also, by specifying the state of the application with its integrity invariants, Why3 can verify if an operation can break those invariants, given its implementation. In the cases where Why3 is not able to prove an assertion in the program, the framework can present a counter-example~\cite{DBLP:journals/jlp/DaillerHMM18}. That said, it is possible to state that Why3 is capable of performing the safety analysis without requiring any changes within the framework.

\subsection{Commutativity and Stability analysis}
\label{comm_analysis}

Having in mind the generic program that was presented in Figure~\ref{generic_prog}, our tool generates the function in Figure \ref{comm} for the commutativity and stability analysis between the operations~\of{f}~and~\of{g}. Our plug-in was implemented with the purpose of automatically generating functions that verify the commutativity and stability between pairs of operations. Given the code from the application, CISE3 uses Why3's parser to obtain an in memory representation of the contents of a WhyML program. After the program is parsed, for each pair of different operations a commutativity analysis function is generated automatically, which also verifies the stability between those operations. If all the generated verification condition for the function in Figure \ref{comm} are discharged then the operations involved commute and do not conflict. This function starts by generating two equal application states, as well as the arguments for each operation so that the preconditions of the analysed operations are preserved. After that, operations \of{f} and \of{g} are executed over state \of{state1}, following a specific order, and after that they are executed in the alternative order over state \of{state2}. Following both executions, if the resulting final states are the same then operations \of{f} and \of{g} commute.

\begin{figure}
	\begin{why3}
	let ghost f_g_commutativity () : ($\tau$, $\tau$)
	 ensures { match result with x1, x2 -> x1 == x2 end }
  	= val $\overline{\mathtt{x_1: \tau_1}}$ in
    	 val state$_1$ : $\tau$ in
    	 val $\overline{\mathtt{x_2: \tau_2}}$ in
    	 val state$_2$ : $\tau$ in
    	 assume { $\mathcal{P}_1 \wedge \mathcal{P}_2 \; \wedge $ state$_1$ == state$_2$ }
    	 f $\overline{\mathtt{x}_1}$ state$_1$;
    	 g $\overline{\mathtt{x}_2}$ state$_1$;
     	 g $\overline{\mathtt{x}_2}$ state$_2$;
    	 f $\overline{\mathtt{x}_1}$ state$_2$;
    	 (state$_1$, state$_2$)
	\end{why3}
	\caption{Generated commutativity analysis function.}
	\label{comm}
\end{figure}

Regarding the stability analysis between \of{f} and \of{g}, if our tool is not able to prove every precondition of operation \of{f} when preceded by the execution of operation \of{g}, then we assume that they are conflicting and they cannot be safely executed concurrently. Operations \of{f} and \of{g} are also considered conflicting in case one cannot prove every precondition of operation \of{g} when preceded by the execution of \of{f}. Finally, if we can prove every assertion of the generated function with our tool, then operations \of{f} and \of{g} are not conflicting. This way, in a single step, we can perform the commutativity and stability analysis for each pair of different operations.

In the postcondition of the function in Figure \ref{comm}, there is a state equality relation (\of{==}). This equality relation is a point-wise field comparison of the record of the state of the application. The equality relation between states can be automatically generated by our tool, if the programmer does not provide one. In Section \ref{case} we present an example that requires an equality relation provided by the programmer.

Finally, for the remaining stability analysis, our tool generates a function for each operation in order to analyse the possibility of multiple concurrent executions of one operation. So, having in mind the generic program presented in Figure \ref{generic_prog}, our tool generates a function for the stability analysis of operation \of{f}, as presented in Figure~\ref{stability}.

\begin{figure}
	\begin{why3}
  		let ghost f_stability () : unit
  		= val $\overline{\mathtt{x_1: \tau_1}}$ in
    		   val state$_1$ : $\tau$ in
   		   val $\overline{\mathtt{x_2: \tau_1}}$ in
   		   val state$_2$ : $\tau$ in
   		   assume { $\mathcal{P}_1  \; \wedge $ state$_1$ == state$_2$ }
   		  f $\overline{\mathtt{x}_1}$ state$_1$;
   		  f $\overline{\mathtt{x}_2}$ state$_1$;
	\end{why3}
	\caption{Generated stability analysis function.}
	\label{stability}
\end{figure}

The function presented in Figure \ref{stability}, starts by generating the initial state of the application and the arguments for the operation calls, so that the preconditions of the first call to operation \of{f} are preserved. Lastly, operation \of{f} is executed consecutively. For the programmer to know if an operation is not conflicting with itself, we use our tool to prove the generated function for the stability analysis of that operation. We assume \of{f} is cannot be safely executed in concurrence with itself if the preconditions of the second call to \of{f} are not preserved.

\subsection{Token System}
\label{tokens}

After our tool's analyses, the programmer is aware of the pairs of operations
that cannot safely execute concurrently. With this information in hand, the
programmer can provide a token system to our tool and check if the consistency
model that it represents is sound, by executing the tool over the application's
specification again. Given the token system, if our tool is able to prove every
generated verification condition, then the specified consistency model is
considered sound. In Section \ref{case} we show the definition of token systems
for our case study, and its repercussions in the performed analyses. For the
specification of token systems, we provide the BNF of the token definition
language in Figure \ref{tkn_lang}.

\begin{figure}
\begin{centering}
	$$
	\begin{array}{rrl}
	\mathit{tokenSystem} & ::= & \mathit{tokenDef} \\
	& | & \mathit{conflictsDef} \\
	\mathit{tokensDef} & ::= & \mathit{token} \: \mathit{tokensDef} \\
	& | & token \\
	\mathit{token} & ::= & \textbf{token} \: \mathit{opId} \:
                               \mathit{tokenId}^+ \\
	& | & \textbf{argtoken} \: \mathit{opId} \: \mathit{argId} \:
              \mathit{tokenId} \\
	\mathit{conflictsDef} & ::= & \mathit{conflict} \: \mathit{conflictsDef}
          \\
	& | & \mathit{conflict} \\
	\mathit{conflict} & ::= & \mathit{tokenId} \: \textbf{conflicts} \:
                                  \mathit{tokenId}
	\end{array}
	$$
	\end{centering}
	\caption{BNF of the token system specification language}
	\label{tkn_lang}
\end{figure}

The first rule from the $token$ production, describes the declaration of a
non-empty list of tokens, associated with an operation. Each token can only be
declared once, as it cannot be associated to more than one operation. The second
rule from the $token$ production describes the association of tokens to
arguments of an operation. The last production, illustrates how the programmer
can declare two tokens as being in conflict. The tokens that are used in the
$\mathit{conflict}$ production must both have been defined previously. When two
tokens declared with the keyword $\textbf{token}$ are conflicting, the
operations associated to those tokens cannot be executed concurrently in any
situation. In this case, our tool will not generate any analysis function
regarding the analysis for the pair of conflicting operations. If two tokens
that were declared with the $\textbf{argtoken}$ keyword are conflicting, the
operations associated to the tokens can only execute concurrently if the value
of the arguments are different. In this case, our tool assumes that two
operations with conflicting arguments only execute concurrently when the values
of the arguments are different.

To illustrate our token system specification language let us consider the following example: our tool is executed over the application from Figure \ref{generic_prog} and finds out operation \of{f} is conflicting with itself. Given this information, the programmer can provide the following token system:

\begin{why3}
		token f t1
		t1 conflicts t1
\end{why3}

The token system presented above shows a consistency model
where operation \of{f} cannot be safely executed
in concurrence with itself in any situation. So, given this token system our tool does not generate the stability analysis \of{f_stability}. However, let us consider that the conflict regarding operation \of{f} is only related to one of its parameters \of{arg}. In this case the programmer can provide a more refined token system like the one below:

\begin{why3}
		argtoken f arg t1
		t1 conflicts t1
\end{why3}

Now this token system depicts a consistency model where \of{f} can only be
safely executed concurrently when the value of \of{arg} is different in each
concurrent execution of \of{f}. Given this new token system, our tool changes
the \of{assume} expression from the \of{f_stability} function in Figure
\ref{stability}. The new \of{assume} expression is changed as follows:

	\begin{why3}
 		assume { $\mathcal{P}_1  \; \wedge \;\mathtt{arg_1} <> \mathtt{arg_2} \; \wedge$ state$_1$ == state$_2$ }
	\end{why3}

\subsection{Strongest postcondition calculus}
\label{sp_calculus}

In general, postconditions are more difficult to write than preconditions. So,
in our tool we introduce a strongest postcondition generator. The strongest
postcondition is a predicate transformer used to automatically generate the
unique strongest postcondition $\mathcal{Q}$ of a function $\mathbf{S}$ and its
preconditions $\mathcal{P}$, in a way that satisfies the Hoare's triple
\{$\mathcal{P}$\} $\mathcal{S}$ \{$\mathcal{Q}$\}
\cite{post,Hoare:1969:ABC:363235.363259}. By using a predicate transformer the
programmer has less of a burden when it comes to the specification effort, which
is our goal.

The chosen target programming language used for the strongest postcondition
calculus, is a similar representation of a subset of the WhyML language. Since
the DSL we chose is similar to a subset of the WhyML language, it is easy to
integrate with our tool. The programmer only needs to provide as input a file
with the specification of a function and its preconditions written with the
DSL. Given that input, our strongest postcondition predicate transformer
automatically generates the postcondition for the specified function. The
generated strongest postcondition gives the programmer a clear hint about which
postcondition she must include in the specification.

\section{CRDTs library}
\label{crdts}

One issue that can occur in geo-replicated systems is the divergence of its data
due to the execution of non-commutative operations in different orders, in
different replicas. A possible solution to this issue is the inclusion of
Conflict-free Replicated Data Types (CRDTs) \cite{CRDT_preguica}. CRDTs are
mutable objects replicated over a set of replicas interconnected by an
asynchronous network. These data types guarantee convergence in a
self-established way, despite failures in the network. A client of a CRDT object
has two types of operations that can be called in order to communicate with it:
\textit{read} and \textit{update}.

To enrich our work, we implement a library of verified CRDTs using Why3. The
goal of this library is to provide a collection of off-the-shelf CRDTs that the
programmer can use in order to solve commutativity issues found in
applications. An example of a CRDT specification from our library is the
Remove-Wins Set presented in Figure \ref{rem-wins}. In this specification a set is represented by two sets: \of{remove_wins_add} stores the elements that were added and \of{remove_wins_removes} stores the elements that have been deleted. In order for an element to be considered as belonging to the set, it needs to be stored in \of{remove_wins_add} not be stored in \of{remove_wins_removes}. This means that when an element is removed from the set, then it cannot be inserted ever again.

\begin{figure}
	\begin{why3}
  type remove_wins_set 'a = {
    mutable remove_wins_add: fset 'a;
    mutable remove_wins_removes: fset 'a; }
		
  let ghost predicate equal (s1 s2: remove_wins_set 'a)
  = s1.remove_wins_add     == s2.remove_wins_add &&
    s1.remove_wins_removes == s2.remove_wins_removes
		
  val empty_set () : remove_wins_set 'a
    ensures { is_empty result.remove_wins_add }
    ensures { is_empty result.remove_wins_removes }
		
  predicate in_set (elt: 'a) (s: remove_wins_set 'a)
  = mem elt s.remove_wins_add &&
    not (mem elt s.remove_wins_removes)
		
  val add_element (elt: 'a) (s: remove_wins_set 'a) : unit
    writes  { s.remove_wins_add }
    ensures { s.remove_wins_add = 
    	      add elt (old s).remove_wins_add }
		
  val remove_element (elt: 'a) (s: remove_wins_set 'a): unit
    writes  { s.remove_wins_removes }
    ensures { s.remove_wins_removes =
              add elt (old s).remove_wins_removes }
	\end{why3}
	\caption{Remove-wins Set CRDT specification in Why3}
	\label{rem-wins}
\end{figure}

\section{Case study}
\label{case}

In this section we present a complete case study that illustrates how our tool
works. In this case study we have a school registration system where we have a
set of students, a set of courses, and an enrollment relation between students
and courses. The implementation and specification for this case study, are
presented in Figure~\ref{school_spec}.

\begin{figure}
	\begin{why3}
type state [@state] = {
  mutable students : fset int;
  mutable courses  : fset int;
  mutable enrolled : fset (int,int);
} invariant{ forall i,j. mem (i,j) enrolled ->
   mem i students /\ mem j courses }

let ghost addCourse (course : int) (state : state): unit
  requires { course > 0 }
  ensures  { state.courses =
             add course (old state).courses}
= state.courses <- add course state.courses

let ghost addStudent (student : int) (state : state): unit
  requires { student > 0 }
  ensures  { state.students =
             add student (old state).students }
= state.students <- add student state.students

let ghost enroll (student course : int) (state : state): unit
  requires { student > 0 /\ course > 0 }
  requires { mem student state.students }
  requires { mem course state.courses }
  requires { not (mem (student,course)
              state.enrolled.remove_wins_removes) }
  ensures  { state.enrolled =
             add (student,course) (old state).enrolled }
= state.enrolled <- add (student,course) state.enrolled

let ghost remCourse (course : int) (state : state): unit
  requires { course > 0 }
  requires { forall i.
               not (mem (i, course) state.enrolled) }
  requires { mem course state.courses}
  ensures  { not (mem course state.courses) }
  ensures  { forall c. c <> course ->
               mem c (old state).courses <->
               mem c state.courses }
= state.courses <- remove course state.courses

predicate state_equality [@state_eq] (s1 s2 : state) =
  s1.students == s2.students &&
  s1.courses  == s2.courses  &&
  s1.enrolled == s2.enrolled
	\end{why3}
	\caption{Specification and implementation of school registration system.}
	\label{school_spec}
\end{figure}

Every field of the state of the application in Figure~\ref{school_spec} is
represented as an \of{fset} which is a finite set from Why3. The invariant
associated to the state of the application declares that a student can only be
enrolled in an existing course. Function \of{mem} from the Why3 standard library represents an ownership relation for sets.

In this case study, operation \of{addCourse} adds a new course to the system,
\of{addStudent} adds a new student to the system, \of{enroll} registers a
student in a course, and lastly \of{remCourse} removes a course from the
system. The specification of operation \of{enroll} states that the student and
course must exist in the system, and, after its execution, the student is
enrolled in the course. The specification of \of{remCourse} indicates that it is
required that no student is enrolled in the course being removed and that the
course must exist in the system. After its execution it is ensured that the
course no longer exists in the system.

To illustrate our strongest postcondition calculus, we present its execution
over operation \of{addCourse}. To execute our predicate transformer the
programmer needs to provide as an input to our tool, the specification presented below: 

\begin{why3}
  addCourse (course : int) (state : state): unit
    requires { course > 0 }
  = state.courses <- add (course, state.courses)
\end{why3}

Over that specification our tool applies the predicate
transformer which generates the following strongest postcondition: 

\begin{why3}
  exists v0. state.courses = add (course,v0) &&
         course > 0
\end{why3}

The generated postcondition states that exists a
certain~\texttt{v0} to which a non-negative $\mathtt{course}$ was
added. Comparing with the postcondition of \of{addCourse} from Figure
\ref{school_spec}, our generated postcondition is more verbose however, this is automatically generated. This helps the programmer to understand which postcondition must be supplied. In fact, an appropriate witness for the existentially quantified variable~\texttt{v0} is \of{(old state).courses}, in
which we recover the postcondition supplied in Figure~\ref{school_spec}.

A particularity of this case study is the introduction of the
\of{state_equality} predicate, which represents the equality relation between
states and is identified by the tag \of{[@state_eq]}. Without the predicate
\of{state_equality}, our tool would generate an equality relation, as we saw in
Section~\ref{cise3}. In the specific case of set comparison, a simple structural
comparison would not be enough to prove if two sets are equal. In order to prove
the equality between sets we need an extensional equality relation over sets,
hence the need for the programmer to specify the \of{state_equality} predicate.

In the implementation presented in Figure~\ref{school_spec}, we can see that the
application has four operations: \of{addCourse}, \of{addStudent}, \of{enroll},
and \of{remCourse}. Given this, our tool generates six different functions for
the commutativity analysis, but we only show one of the generated functions in
Figure~\ref{school_comm}.

\begin{figure}
	\begin{why3}
let ghost enroll_remCourse_commutativity () : (state, state)
  ensures  { match result with
               x1, x2 -> state_equality x1 x2
             end }
= let ghost student1 = any int in
  let ghost course1 = any int in
  let ghost state1 = any state in
  let ghost course2 = any int in
  let ghost state2 = any state in
  assume { (student1 > 0 /\ course1 > 0) /\
    mem student1 (students state1) /\
    mem course1 (courses state1) /\
    not mem ((student1, course1)
    (enrolled state1)) /\
    course2 > 0 /\
    (forall i. not mem (i, course2) (enrolled state2)) /\
    mem course2  (courses state2) /\
    state_equality state1 state2 };
  remCourse course2 state1;
  enroll student1 course1 state1;
  enroll student1 course1 state2;
  remCourse course2 state2;
  (state1, state2)
	\end{why3}
	\caption{Commutativity analysis function for operations \of{enroll} and \of{remCourse}.}
	\label{school_comm}
\end{figure}

The function presented in Figure \ref{school_comm}, starts by generating the
arguments used to call the analysed operations. The \of{assume} expression is
used to restrict the state for the generated arguments, so that the
preconditions are preserved and both generated states are equal. After that,
operation \of{remCourse} is called, followed by a call to operation \of{enroll}
over the state \of{state1}. If Why3 is not able to prove the preconditions for
the call of operation \of{enroll} over \of{state1}, then operations \of{enroll}
and \of{remCourse} cannot be safely executed concurrently. Next, operation
\of{enroll} is called over state \of{state2} followed by a call to operation
\of{remCourse} also over state \of{state2}. In a similar way that Why3 verified
the preconditions of operation \of{enroll} in the previous order of execution,
in this order we will observe if Why3 can prove the preconditions of the call to
\of{remCourse}. If these preconditions are not preserved, then \of{enroll} and
\of{remCourse} are conflicting and cannot be safely executed
concurrently. Lastly, the pair \of{(state1, state2)} is returned, and if the
elements are equal we assume that operations \of{enroll} and \of{remCourse}
commute. In the case of the function shown in Figure \ref{school_comm} we are
not able to prove the preservation of the preconditions of operation \of{enroll}
when executed after \of{remCourse}. Consequently, the equality between states
after the execution of both operations in alternative orders cannot be
proved. Thus we conclude that operations \of{enroll} and \of{remCourse} conflict
and we cannot prove that they commute.

Now we proceed to the remaining stability analysis where we check if multiple
concurrent executions of the same operation can occur concurrently. To do so,
our tool generates a stability analysis function for each operation as specified
in Section \ref{comm_analysis}. As an example, in Figure \ref{school_stability}
we present the generated function for the stability analysis of operation
\of{remCourse}. Initially, each stability analysis function generates the
arguments used in the calls to the operation. As in the commutativity analysis,
the \of{assume} expression restricts the space of possible combinations of
values for the generated arguments. Then, we call the operation being analysed
consecutively and if Why3 is not able to prove the preservation of the
preconditions \textit{a priori} of the second call to the operation, then we
assume the operation is conflicting with itself. In the case of the function
presented in Figure \ref{school_stability}, every verification condition
generated for the function \of{remCourse_stability} is proved automatically,
which allows us to conclude that the operation \of{remCourse} can be executed
safely concurrently with itself.

\begin{figure}
	\begin{why3}
  let ghost remCourse_stability () : ()
  = let ghost course1 = any int in
    let ghost state1 = any state in
    let ghost course2 = any int in
    let ghost state2 = any state in
    assume { (course1 > 0 /\
      (forall i. not mem (i, course1) (enrolled state1)) /\
      mem course1 (courses state1) /\
      course2 > 0 /\
      (forall i. not mem (i, course2) (enrolled state2)) /\
      mem course2 (courses state2) /\
      state_equality state1 state2 };
    remCourse course1 state1;
    remCourse course2 state1
	\end{why3}
	\caption{Stability analysis function for the school registration system.}
	\label{school_stability}
\end{figure}

As we stated before, in this application the operations \of{enroll} and
\of{remCourse} are conflicting. Having this in mind and resorting to the language presented in Section \ref{tokens} we can define a token system for the application and verify its soundness. One possible token system that the programmer can specify is the following:

\begin{why3}
		token enroll t1
		token remCourse t2
		t1 conflicts t2
\end{why3}

This token system represents a consistency model where the conflicting
operations \of{enroll} and \of{remCourse} cannot be executed concurrently in any
situation. If our tool analysed this token system then it would not generate the
function regarding the commutativity and stability analysis for operations
\of{enroll} and \of{remCourse}. This way our tool is able to prove every
generated verification condition, meaning that the specified consistency model
is sound. However, this model is too strict because if we analyse the conflict
between \of{enroll} and \of{remCourse} more closely, we understand that the
conflict depends on the argument \of{course}. So, it is possible to define a
more refined consistency model represented by the token system presented below:

\begin{why3}
		argtoken enrol course t1
 		argtoken remCourse course t2
		t1 conflicts t2
\end{why3}

The above token system states that the argument \of{course} of operation
\of{enroll}, associated to token \of{t1} is conflicting with the argument
\of{course} of operation \of{remCourse} which is associated to token \of{t2}. So
this token system depicts a consistency model where operations \of{enroll} and
\of{remCourse}, can only be executed concurrently for different course
arguments. That said, we need to modify the \of{assume} expression from
operation \of{enroll_remCourse_commutativity}, adding the restriction that the
argument \of{course} must have different values in each concurrent execution of
the operations \of{enroll} and \of{remCourse}. With this modification Why3 is
able to prove that these operations are stable implying that the consistency
model specified in the second given token system is sound.

Lastly, in this case study operations \of{addCourse} and \of{remCourse} do not commute. To solve this issue, the programmer can replace the data structure responsible for storing courses by a Remove-Wins CRDT Set. Due to this modification, the state of the application is changed to the one  seen in Figure \ref{school_crdt}.

\begin{figure}
	\begin{why3}
  type state [@state] = {
    mutable students : fset int;
    mutable courses  : remove_wins_set int;
    mutable enrolled : fset (int,int);
  } invariant { forall i, j. mem (i,j) enrolled ->
    mem i students /\ in_set j courses }
	\end{why3}
	\caption{School registration system state with a CRDT}
	\label{school_crdt}
\end{figure}

Apart from the changes made to the state of the application, the programmer also
needs to modify the operations that manipulate the collection of courses, in
order to respect the CRDT's API. The introduction of this CRDT solves this
commutativity issue because conflicting concurrent updates are solved according
to a deterministic conflict resolution policy.

% now when \of{addCourse} is executed, at the end of its execution the course is
% only guaranteed to be in the \of{remove_wins_add} field from \of{courses}
% Remove-Wins set. This guarantee is not enough to state that the added course
% exists in the system as we saw in Section~\ref{crdts}. However, when operation
% \of{remCourse} is executed the course is guaranteed to be out of the
% system. This happens because in our implementation of the Remove-Wins set, we
% give precedence to operation \of{remCourse}. So now, independent of the order
% of execution of \of{remCourse} and \of{addCourse}, our CRDT will guarantee
% that the system converges.

\section{Related Work}
\label{related}

In this section we present an overview of static analysis tools similar to the
one we present in this paper, in the sense that they are all used to reason
about the verification of distributed applications.

Quelea is a tool used to verify distributed systems built over eventually
consistent replicated databases~\cite{quelea}. The approach of this tool is
based on a contract language that allows for the specification of fine-grained
properties about the consistency of an application. The verification conditions
derived from these contracts are then proved using Z3. For each operation, this
tool verifies which consistency model is the most appropriate, \textit{i.e.},
which consistency model has its restrictions satisfied by the operation's
contract. The complexity of the specification of these contracts is high, since
it makes the programmer reason about possible concurrent
interferences. Additionally, there is no guarantee that the contracts are
sufficient to assure the preservation of the application's invariants.

Q9 is another tool that analyses applications built over replicated databases
that use eventual consistency~\cite{Kaki:2018:SRT:3288538.3276534}. Q9 discovers
anomalies in the correction of the application using a bounded verification
technique, and solves them automatically. The bounded verification technique
analyses a search space that is restricted by the number of concurrent effects
that can occur over the state of the application. Since there is a restriction
over the maximum number of concurrent effects that can occur to the state of the
application, this tool is not capable of assuring the complete correction of the
system. Our tool does not restrict the number of concurrent effects that can
occur over the state, whereas Q9 does.

Repliss is a tool that verifies applications that execute over weakly consistent
databases, given their specification, the integrity invariants and the
implementation of the
application~\cite{Zeller:2017:TPW:3064889.3064893}. Repliss provides a DSL that
is used by the programmer to write the required input. With the input
provided by the programmer, Repliss translates the program into a sequential
Why3 program. If the sequential program is proved correct, then the
initial program is also considered correct. In comparison with our
tool, the DSL presented in Repliss is more limited than the WhyML language. So,
by developing one application, for our tool to analyse, directly over the WhyML language, it is easier to specify its behaviour.

The Hamsaz framework given the specification of the system and using CVC4,
determines the conflicts and causal dependencies between operations~\cite{Hamsaz}. For the specification one needs to define an object that
includes the type of the state, the integrity invariants and the methods of the
application. The goal of this tool is to automatically obtain a replicated
system that is correct by-construction, that
converges and preserves data integrity. Simultaneously, the obtained system avoids unnecessary
synchronisation having in mind the conflict and dependency relations between
operations.

One advantage our tool has over the aforementioned tools, is the inclusion of a
strongest postcondition predicate transformer. With this the programmer does not
have to reason about the specification of postconditions.

\section{Conclusion}
\label{conclusion}

In this paper we explored an automatic approach for the analysis of weakly
consistent applications using the deductive verification framework Why3. We
propose that the programmer provides a sequential specification and
implementation. After that, the programmer uses CISE3 to reason about the pairs
of conflicting and causally dependent operations from the application. With this
information the programmer can then use a CRDT to solve commutativity issues,
and specify a token system in order to assess if a specific consistency model is
sound over the application. In order to aid the programmer's task, our tool
features a strongest postcondition predicate generator. Our proposal is
illustrated in Section~\ref{case} by means of a case study implemented and
verified with our tool. Besides the presented case study, we have also
successfully verified several other case studies, such as a bank application as
well as an auction application.

The next steps related to our work are the following: improve and expand the
library of CRDTs and refine our strongest postcondition predicate transformer.

Currently we have a library of CRDTs that only has a few simple implementations,
like the one in Appendix \ref{rem-wins}. Thus, our goal is to optimize the
already existing CRDT implementations, and add new ones so our library can be
used in a wider variety of examples.

Also, the target language we provide for the strongest postcondition predicate
transformer has a few limitations, like the lack of possibility to specify
loops. So, our goal regarding this feature is to refine said language and add
more features to it. One of the features we want to implement is the support of
[for ... each] constructors, since mainly this is the kind of loops that are
used in applications that operate over replicated databases. By improving our
target language, we can also expand our strongest postcondition predicate
transformer. This way, this language can be used for the generation of
postconditions for an ampler set of applications.

\bibliographystyle{plain}
\bibliography{bibliography}

\end{document}